\documentclass[a4paper,11pt]{article}
\pdfoutput=1 

\usepackage{jheppub} 

\usepackage[T1]{fontenc} 
\usepackage[textsize=footnotesize,textwidth=2.5cm]{todo notes}

\definecolor{mikadoyellow}{rgb} {0.16, 0.254, 0.6}

\usepackage{amssymb,amsmath,latexsym,bm,amsfonts}
\usepackage{graphicx}
\usepackage{longtable}
\usepackage{color,xcolor}
\usepackage{indentfirst}
\usepackage{subfigure}
 \usepackage{comment}

\title{ \boldmath Formulas for Partial Entanglement Entropy}

\author[a,b]{Qiang Wen}

\affiliation[a]{Shing-Tung Yau Center of Southeast University, Nanjing 210096, China}
\affiliation[b]{School of Mathematics, Southeast University, Nanjing 211189, China}

\emailAdd{wenqiang@seu.edu.cn}

\abstract{The partial entanglement entropy (PEE) $s_{\mathcal{A}}(\mathcal{A}_i)$ characterizes how much the subset $\mathcal{A}_i$ of $\mathcal{A}$ contribute to the entanglement entropy $S_{\mathcal{A}}$. We find one additional physical requirement for $s_{\mathcal{A}}(\mathcal{A}_i)$, which is the invariance under a permutation. The partial entanglement entropy proposal satisfies all the physical requirements. We show that for Poincar\'e invariant theories the physical requirements are enough to uniquely determine the PEE (or the entanglement contour) to satisfy a general formula. This is the first time we find the PEE can be uniquely determined. Since the solution of the requirements is unique and the \textit{PEE proposal} is a solution, the \textit{PEE proposal} is justified for Poincar\'e invariant theories.}

\begin{document} 
\maketitle
\flushbottom
\section{Introduction}
Entanglement entropy $S_{\mathcal{A}}$, which characterizes the correlation between a region $\mathcal{A}$ and its complement $\bar{\mathcal{A}}$ is the most important quantity that we have used to explore the entanglement structure of a quantum system. Nevertheless, entanglement entropy in quantum field theory is an ambiguous quantity. Because of the short distance correlation, entanglement entropy in quantum field theory is infinite thus needs to be regularized properly. The regularization is to ignore certain types of correlations, which can be done by introducing certain types of cutoffs. Nevertheless, different cutoffs mean different ways to count entanglement thus lead to different values for entanglement entropy. Even with the scale cutoff settled down, the typical size fluctuations of the region still make the sub-leading contributions to the entanglement entropy ambiguous \cite{Casini:2009sr,Casini:2015woa}. Due to these ambiguities, people turn to the mutual information defined by
\begin{align}\label{dmutual}
I(A,B)=S_{A}+S_{B}-S_{A\cup B}\,.
\end{align}
For any two non-intersecting regions $A$ and $B$, $I(A,B)$ is finite and cutoff independent but still capture the information of entanglement.

Recently, several papers \cite{Vidal:2002rm,Botero,Vidal,PhysRevB.92.115129,Coser:2017dtb,Tonni:2017jom,Alba:2018ime,Wen:2018whg,Wen:2018mev,Kudler-Flam:2019oru,Wen:2019ubu,DiGiulio:2019lpb,Han:2019scu,Ageev:2019fjf,Kudler-Flam:2019nhr,Roy:2019gbi,deBuruaga:2019xwv,MacCormack:2020auw} studied the so-called entanglement contour \cite{Vidal}, which is a function that characterizes how much each degrees of freedom in a region $\mathcal{A}$ contributes to the entanglement entropy $S_{\mathcal{A}}$. In other words, consider a quantum field theory in $d$ dimensions (in this paper $d$ means the dimension of spacetime), the entanglement contour is a density function of entanglement entropy that depends on $\mathcal{A}$ and satisfies
\begin{align}
S_{\mathcal{A}}=\int_{ \mathcal{A}}f_{\mathcal{A}}(\textbf{x})d\sigma_{\textbf{x}}\,.
\end{align}
where $\textbf{x}$ denotes a point in $\mathcal{A}$ and $\sigma_{\textbf{x}}$ denotes a infinitesimal subset of $\mathcal{A}$ at $\textbf{x}$.
It is more convenient to study the \textit{partial entanglement entropy} (PEE) $s_{\mathcal{A}}(\mathcal{A}_i)$ for any subset $\mathcal{A}_i$ of $\mathcal{A}$, which is defined in the following way
\begin{align}\label{a2definition}
s_{\mathcal{A}}(\mathcal{A}_i)=\int_{ \mathcal{A}_i}f_{\mathcal{A}}(\textbf{x})d\sigma_{\textbf{x}}\,.
\end{align}
In other words $s_{\mathcal{A}}(\mathcal{A}_i)$ captures the contribution from $\mathcal{A}_i$ to the entanglement entropy $S_{\mathcal{A}}$. Like the mutual information, the PEE is finite and cutoff independent when the boundaries of $\mathcal{A}$ and $\mathcal{A}_i$ donot overlap. The PEE explores the local properties of quantum entanglement. 

So far the fundamental definition of the PEE (or entanglement contour) based on the reduced density matrix is still not clear. If the PEE can be well defined, it should satisfy the following requirements \cite{Vidal}:
\begin{enumerate}
\item \textbf{Additivity}: by definition we should have
\begin{align}\label{additivity}
s_{\mathcal{A}}(\mathcal{A}_{i})&=s_{\mathcal{A}}(\mathcal{A}_{i}^{a})+s_{\mathcal{A}}(\mathcal{A}_{i}^{b})\,, \quad \mathcal{A}_{i}=\mathcal{A}_{i}^{a}\cup \mathcal{A}_{i}^{b}.
\end{align}

\item \textbf{Invariance under local unitary transformations}: $s_{\mathcal{A}}(\mathcal{A}_i)$ is invariant under any local unitary transformations act only inside $\mathcal{A}_i$ and $\bar{\mathcal{A}}$.

\item \textbf{Symmetry}: For any symmetry transformation $\mathcal{T}$ under which $\mathcal{T}\mathcal{A}=\mathcal{A}'$ and  $\mathcal{T} \mathcal{A}_{i}= \mathcal{A}'_{i}$, we have $s_{\mathcal{A}}(\mathcal{A}_i)=s_{\mathcal{A}'}(\mathcal{A}'_i)$.

\item \textbf{Normalization}: $ S_{\mathcal{A}}=s_{\mathcal{A}}(\mathcal{A}_{i})|_{\mathcal{A}_i\to\mathcal{A}}\,.$

\item \textbf{Positivity}: $s_{\mathcal{A}}(\mathcal{A}_i)\ge 0$.

\item \textbf{Upper bound}: $
 s_{\mathcal{A}}(\mathcal{A}_i) \leq S_{\mathcal{A}_{i}} \,.
$
\end{enumerate}

However, the above requirements are not enough to determine the PEE uniquely. So far, there are three proposals to construct the PEE (or entanglement contour) that satisfies the above requirements. Each of them are restricted to some special cases. The first one is the Gaussian formula \cite{Botero,Vidal,PhysRevB.92.115129,Coser:2017dtb,Tonni:2017jom,Alba:2018ime,DiGiulio:2019lpb,Kudler-Flam:2019nhr} that applies to the Gaussian states in free theories that can be completely characterized in terms of the correlation matrix. In these cases, the reduced density matrix can be block diagonalized and a natural probability weight can be assigned to each site in the region. Following this analysis, the entanglement entropy of a region can be written as a summation over all the sites in that region. Nevertheless, it is hard to argue that this summation is the collection of local contributions for entanglement entropy. The second proposal is a geometric construction \cite{Wen:2018whg,Wen:2018mev,Han:2019scu} based on the boundary and bulk modular flows in holographies. It applies to the static spherical regions (or covariant intervals) for holographic field theories. The contour function constructed in this way has a clear physical meaning as the local distribution of entanglement.
The third one is the partial entanglement entropy proposal\footnote{ See also Ref.\cite{Kudler-Flam:2019oru} for its reformulation using conditional entropy, Ref.\cite{Kudler-Flam:2019nhr} for its extension to construct the contour of entanglement negativity, and Ref.\cite{Ageev:2019fjf} for its extension to explore the contour of holographic complexity.} \cite{Wen:2018whg,Wen:2019ubu}  that claims the PEE is given by an additive linear combination of subset entanglement entropies, which we will explicitly discuss in this paper. The consistency check between the \textit{Gaussian formula} and the \textit{PEE proposal} for some cases of free boson and free fermion can be found in Ref. \cite{Kudler-Flam:2019nhr}. Also the analytical results from the fine structure analysis and the \textit{PEE proposal} exactly match with each other \cite{Wen:2018whg,Wen:2018mev,Han:2019scu}. A new way will be introduced in this paper following the construction of extensive (or additive) mutual information (EMI) \cite{Casini:2008wt} (see also Ref. \cite{Roy:2019gbi} for a related construction).

The entanglement contour gives a finer description for the entanglement structure. It allows us to estimate the central charge c of the underlying CFT by studying a single region in $d=2$, and to discriminate between gapped systems and gapless systems with a finite number of zero modes in $d=3$ \cite{Vidal}. It has been shown to be particularly useful to characterize the evolution of the entanglement structure when studying dynamical situations \cite{Vidal,Kudler-Flam:2019oru,DiGiulio:2019lpb}. The local modular flow in $d=2$ can be generated from the PEE in a extremely simple way \cite{Wen:2019ubu}. Also it has been recently demonstrated that the entanglement contour (calculated by the \textit{PEE proposal}) is a useful probe of slowly scrambling and non-thermalizing dynamics for some interacting many-body systems \cite{MacCormack:2020auw}.  The holographic picture of entanglement contour \cite{Wen:2018whg,Wen:2018mev} gives finer correspondence between quantum entanglement and bulk geometry (see Ref. \cite{Abt:2018ywl} for an interesting application). It should be closely related to the other holographic formalisms that attempts to give a finer description for holographic entanglement, such as the tensor network \cite{Swingle:2009bg} and the bit threads picture \cite{Freedman:2016zud} (for related discussions see Refs. \cite{Kudler-Flam:2019oru,Han:2019scu}).  We expect the new concept of entanglement contour in quantum information to play an important role in our understanding of gauge/gravity dualities, entanglement structure of quantum field theories and quantum many-body systems in condensed matter theories.

Since the above requirements are not enough, it is possible that we can find different solutions to those requirements for the same region. However, the PEE or entanglement contour is introduced following a clear physical meaning as a finer description of the underlying entanglement structure of a quantum system, which should be unique. So far, the known contour functions constructed from different proposals are remarkably consistent with each other. So we come to the foundational question in the study of PEE: is the PEE or entanglement contour unique for any state of a quantum system? If it is then is there a unique way to define or determine the PEE? The contour functions we constructed cannot be regarded as the density function of entanglement entropy if entanglement contour is not unique. In this paper, we make progress in answering this question. We point out that the PEE should satisfy another physical requirement, which is a symmetry under permutation. Based on this permutation symmetry and the other known requirements, we follow the discussions \cite{Casini:2008wt} by Casini and Huerta, to find that the physical requirements can give strong enough constraints on PEE. We give an explicit discussion for generic quantum field theories with Poincar\'e symmetry to show that the PEE is unique and should satisfy a general formula. Furthermore, this formula is consistent with the \textit{PEE proposal} and the fine structure analysis. Though the requirement of normalization is very subtle, we show that the \textit{PEE proposal} \cite{Wen:2018whg,Wen:2019ubu} is a solution to all the requirements.

\section{A new physical requirement for partial entanglement entropy}
Since the PEE $s_{\mathcal{A}}(\mathcal{A}_i)$ captures the contribution from the subset $\mathcal{A}_i$ to the entanglement entropy of $\mathcal{A}$, in some sense $s_{\mathcal{A}}(\mathcal{A}_i)$ captures the correlation between $\mathcal{A}_i$ and $\bar{\mathcal{A}}$. Here $\bar{\mathcal{A}}$ is any system that purifies $\mathcal{A}$. Then it will be convenient to write the PEE in the following form
\begin{align}
s_{\mathcal{A}}(\mathcal{A}_i)=\mathcal{I}(\bar{\mathcal{A}},\mathcal{A}_i)\,.
\end{align}
Note that the mutual information is noted as a different symbol $I$. As all the correlations are mutual, it is natural to require $\mathcal{I}(\bar{\mathcal{A}},\mathcal{A}_i)$ to be invariant under the following permutation,
\begin{align}\label{permutation}
\mathcal{I}(\bar{\mathcal{A}},\mathcal{A}_i)=\mathcal{I}(\mathcal{A}_i,\bar{\mathcal{A}})\,.
\end{align} 
In other words we should have $s_{\mathcal{A}}(\mathcal{A}_i)=s_{\bar{\mathcal{A}}_i}(\bar{\mathcal{A}})$. The permutation symmetry together with the requirement of additivity indicate that, $\mathcal{I}(\bar{\mathcal{A}},\mathcal{A}_i)$ can be written as a double integration over $\bar{\mathcal{A}}$ and $\mathcal{A}_i$,
\begin{align}\label{IAB1}
\mathcal{I}(\bar{\mathcal{A}},\mathcal{A}_i)=\int_{\bar{\mathcal{A}}}d\sigma_{\textbf{x}}\int_{\mathcal{A}_i}d\sigma_{\textbf{y}} ~J(\textbf{x},\textbf{y})\,,
\end{align}
where  $\textbf{x}$ ($\textbf{y}$) represents points in $\bar{\mathcal{A}}$ ($\mathcal{A}_i$), $\sigma_{\textbf{x}}$ ($\sigma_{\textbf{y}}$) represents the infinitesimal subset of  $\bar{\mathcal{A}}$ ($\mathcal{A}_i$) at $\textbf{x}$ ($\textbf{y}$). In high energy physics, regions can be covariant (or non-static) in spacetime, hence we should also consider the dependence on the (normal) direction of each infinitesimal subsets in spacetime. The integrand $J(\textbf{x},\textbf{y})$ is just the PEE between the two infinitesimal subsets, i.e.
\begin{align}
J(\textbf{x},\textbf{y})=\mathcal{I}(\sigma_{\textbf{x}},\sigma_{\textbf{y}})\,.
\end{align}

The formula \eqref{IAB1} for the PEE indicates that the PEE between any two non-intersecting regions is the summation over all the PEEs between any pair of degrees of freedom in these two regions. For a discrete system, the PEE can be written as
\begin{align}
\mathcal{I}(\bar{\mathcal{A}},\mathcal{A}_i)=\sum_{i\in \bar{\mathcal{A}}}\sum_{j\in \mathcal{A}_i} J_{ij}\,,
\end{align}
where $J_{ij}$ is the PEE between the $i$th site in $\bar{\mathcal{A}}$ and the $j$th site in $\mathcal{A}_i$.

If we impose the normalization requirement, then it seems that all the entanglement entropies can be generated from the PEE:
\begin{align}\label{IAB2}
S_{\mathcal{A}}=\sum_{i\in \bar{\mathcal{A}}}\sum_{j\in \mathcal{A}} J_{ij}\,.
\end{align}
The above equations have been proposed as a framework \cite{Casini:2008wt,Roy:2019gbi} (see also Ref. \cite{Cao:2016mst})  to evaluate entanglement entropies. In \cite{Roy:2019gbi}, $J_{ij}$ is call the \textit{entanglement adjacency matrix} of the state. If we apply \eqref{IAB2} to disconnected subsets, we find $S_{i\cup j}$, the entanglement entropy of the union of the $i$th and $j$th sites, is given by
\begin{align}
S_{i\cup j}=\sum_{a\neq i,j}J_{a i}+J_{a j}=S_{i}+S_{j}-2 J_{ij}\,.
\end{align}
Then we find the PEE $J_{ij}$ is just given by the half of the mutual information $I(i,j)$ between the $i$th and $j$th site:
\begin{align}
I(i,j)=S_{i}+S_{j}-S_{i\cup j}=2J_{ij}\,.
\end{align}
Also, the mutual information between any two non-intersecting regions $A$ and $B$ is then given by
\begin{align}\label{Jab1}
I(A,B)=2\sum_{i\in A, j\in B}J_{ij}\,,
\end{align}
which is always additive. Though several lattice models\footnote{The examples include  Affleck-Kennedy-Lieb-Tasaki (AKLT) state on a spin-1 chain \cite{Affleck:1987vf}, the \textit{valence bond states} (where qubits are paired into maximally entangled Bell pairs) and the \textit{rainbow chain}.} \cite{Roy:2019gbi} and the massless free fermions \cite{Casini:2008wt} in $d=2$ are shown to have additive mutual information thus the equations \eqref{IAB2} have exact solutions, it is absolutely not true for general cases. Based on the above discussion we may conclude that the requirements 1-6 together with the requirement of the symmetry under permutation \eqref{permutation} are in general not compatible if the normalization requirement is imposed on any regions including the disconnected ones.

Let us consider the simple example of a lattice model on a circle with $N$ sites. The number of all possible subsets is $2^{N}$, so the normalization requirement gives $2^{N}$ equations like \eqref{IAB2}. These equations are usually incompatible because the number of $J_{ij}$ is $N(N-1)/2$, which is much smaller than $2^{N}$ \cite{Roy:2019gbi}. If the solution exists then the entanglement entropies of all the subsets will be highly constraint thus the mutual information is additive.

\section{The partial entanglement entropy proposal as a solution}

\begin{figure}[h] 
   \centering
    \includegraphics[width=0.3 \textwidth]{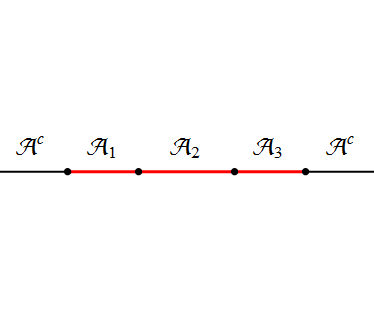}\quad
        \includegraphics[width=0.3
   \textwidth]{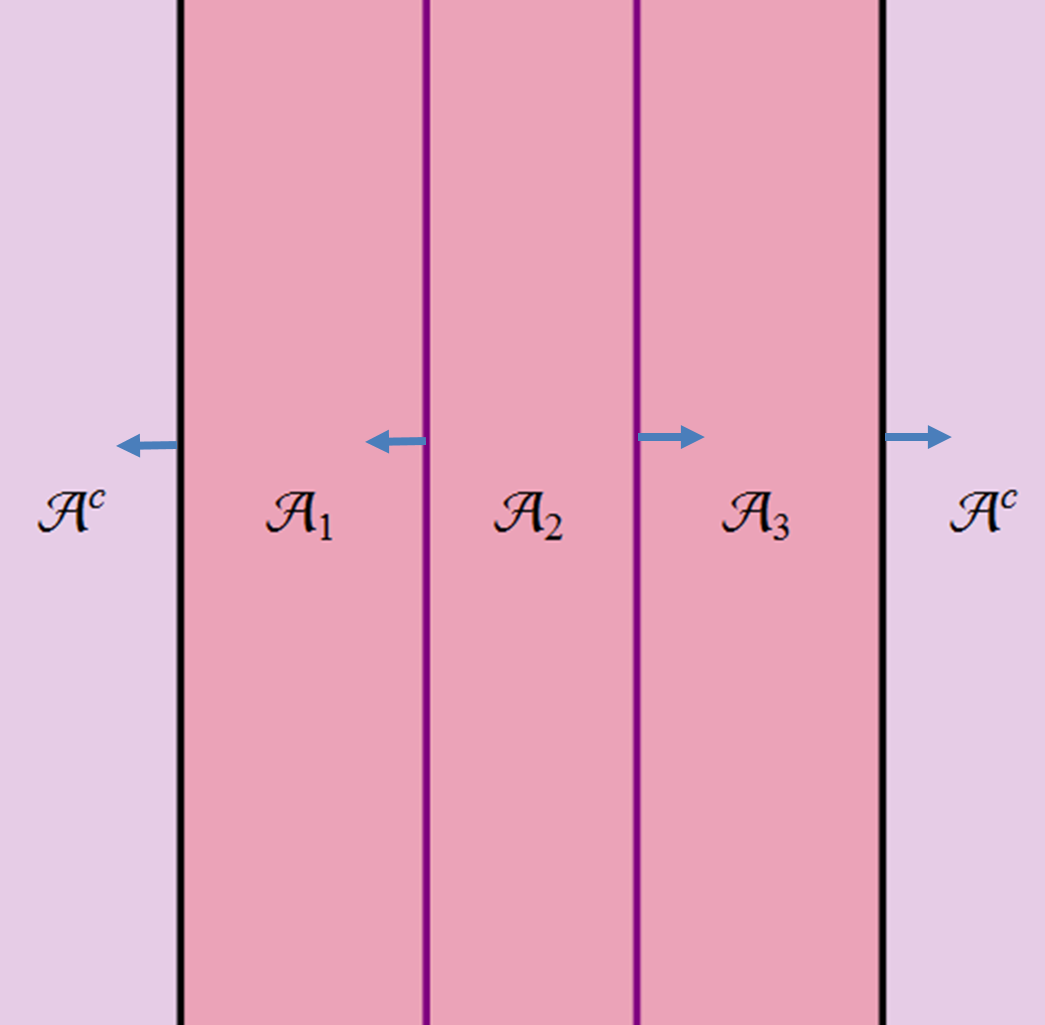}
      \quad\includegraphics[width=0.3 \textwidth]{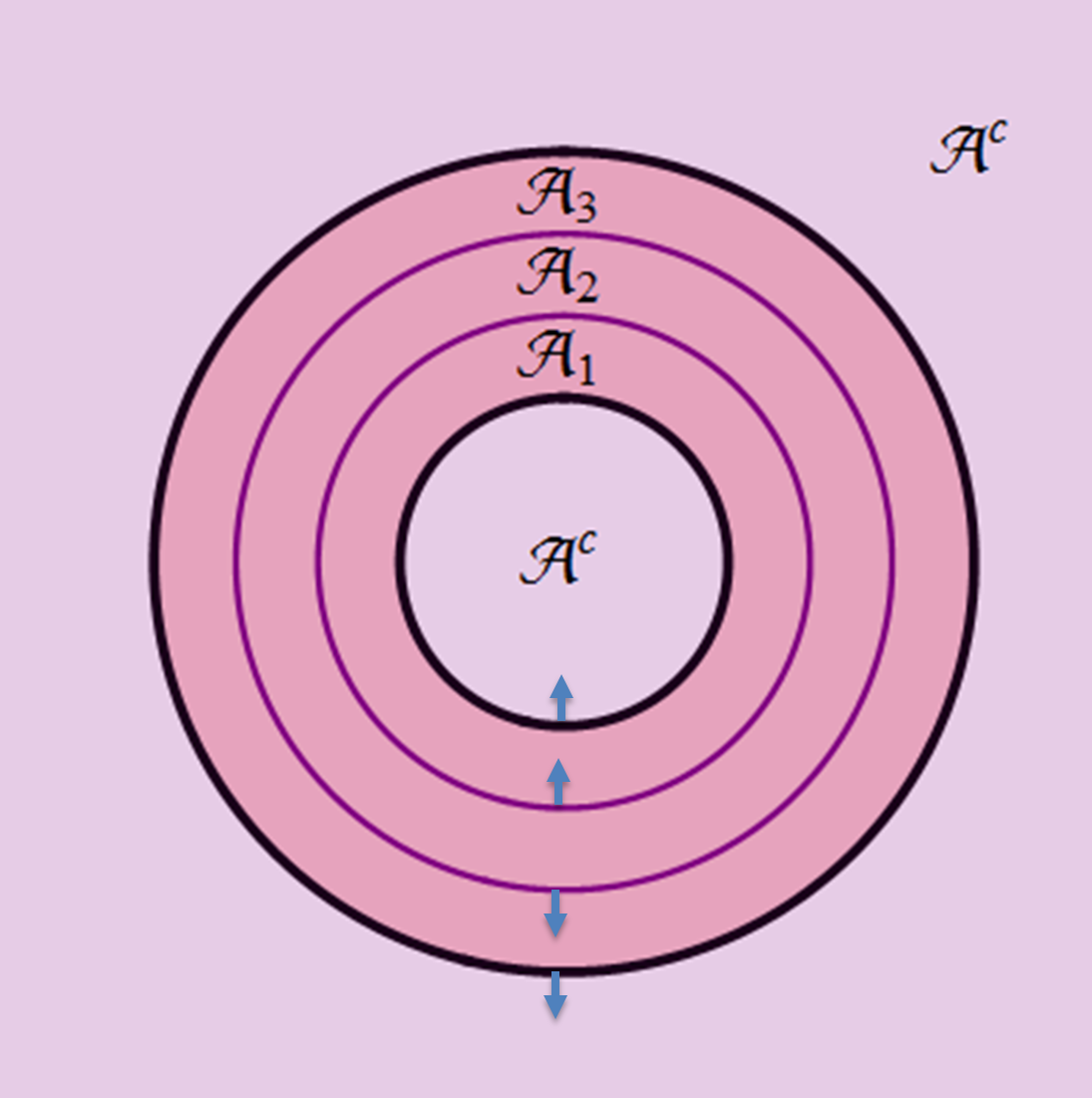} 
 \caption{These figures show examples for the \textit{quasi-one-dimensional} cases, the partitions of an interval, a strip and an annulus. The arrows represent the outward direction of the boundaries $L_{1,2}$ of $\mathcal{A}$ and $l_{1,2}$ of $\mathcal{A}_2$.
\label{1}}
\end{figure}

The \textit{PEE proposal} \cite{Wen:2018whg,Wen:2019ubu} claims that for any two-dimensional theories, the PEE is given by a linear combination of subset entanglement entropies. More explicitly given a connected region $\mathcal{A}$, for any connected subset $\mathcal{A}_2$, which in general divide the region $\mathcal{A}$ into three subsets $\{\mathcal{A}_1,\mathcal{A}_2,\mathcal{A}_3\}$, $s_{\mathcal{A}}(\mathcal{A}_2)$ is given by
\begin{align}\label{ECproposal}
s_{\mathcal{A}}(\mathcal{A}_2)=\frac{1}{2}\left(S_{\mathcal{A}_1\cup\mathcal{A}_2}+S_{\mathcal{A}_2\cup\mathcal{A}_3}-S_{\mathcal{A}_1}-S_{\mathcal{A}_3}\right)\,.
\end{align} 
This proposal can be extended to higher dimensional configurations\footnote{Here the ``configuration'' include the theory, region $\mathcal{A}$ and its partition.} with rotation symmetry or translation symmetry (see Fig.\ref{1}), which we call the \textit{quasi-one-dimensional} configurations. In such cases the contour function respects the symmetries thus only depends on one parameter. Consider the partitions that respect the symmetries, as in the two dimensional cases, the requirement of additivity can be satisfied without imposing extra constraints on the subset entanglement entropies. In Refs.\cite{Wen:2019ubu,Kudler-Flam:2019oru}, it was shown that the proposed PEE \eqref{ECproposal} in \textit{quasi-one-dimensional} configurations satisfies the requirements 1-6 in general theories with no constraints on the subset entanglement entropies. We only need to use the general properties of entanglement entropy such as causality and strong subadditivity. The requirement of normalization is automatically satisfied because when $\mathcal{A}_1$ and $\mathcal{A}_3$ vanish thus $\mathcal{A}_2\to\mathcal{A}$, we find \eqref{ECproposal} recovers $S_{\mathcal{A}}$. Furthermore, assuming $\bar{\mathcal{A}}$ to be a system that purifies the region $\mathcal{A}$, then
\begin{align}
\mathcal{I}(\bar{\mathcal{A}},\mathcal{A}_2)&=s_{\mathcal{A}}(\mathcal{A}_2)
\cr
&=\frac{1}{2}\left(S_{\bar{\mathcal{A}}\cup\mathcal{A}_3}+S_{\bar{\mathcal{A}}\cup\mathcal{A}_1}-S_{\mathcal{A}_1}-S_{\mathcal{A}_3}\right)
\cr
&=S_{\bar{\mathcal{A}}_2}(\bar{\mathcal{A}})=\mathcal{I}(\mathcal{A}_2,\bar{\mathcal{A}})\,,
\end{align}
thus, the requirement of invariance under the permutation \eqref{permutation} is also satisfied. In summary, the PEE \eqref{ECproposal} in \textit{quasi-one-dimensional} configurations is a solution to all the seven physical requirements. This seems to be in contradiction with our previous conclusion that the seven requirements are in general not compatible. Also the PEE \eqref{ECproposal} is definitely not a mutual information.

In order to see this problem more clearly, let us consider again a discrete model and use $\bar{a},a_1,a_2,a_3$ to represent the sites inside $\bar{\mathcal{A}},\mathcal{A}_1,\mathcal{A}_2,\mathcal{A}_3$. For convenience we define 
\begin{align}
 J_{a_i,a_j}=\sum_{i\in a_i,\, j\in a_j}J_{ij}\,.
 \end{align} 
Then we impose the normalization requirement to the subset entanglement entropies and write them as summations of partial entanglement entropies. For example
\begin{align}
S_{\mathcal{A}_1\cup\mathcal{A}_2}=J_{\bar{a},a_1}+J_{\bar{a},a_2}+J_{a_3,a_1}+J_{a_3,a_2}\,.
\end{align}
Then we find
\begin{align}\label{Jab2}
&\frac{1}{2}\left(S_{\mathcal{A}_1\cup\mathcal{A}_2}+S_{\mathcal{A}_2\cup\mathcal{A}_3}-S_{\mathcal{A}_1}-S_{\mathcal{A}_3}\right)=J_{\bar{a},a_2}\,,
\end{align}
which means the \textit{PEE proposal} \eqref{ECproposal} does give the partial entanglement entropy.

We find that if we apply the normalization requirement also for disconnected regions, then half of the EMI \eqref{Jab1} and the PEE \eqref{Jab2} are equivalent to each other. If the normalization requirement only applies to connected regions, then \eqref{Jab2} is the only way to write the PEE as a linear combination of subset entanglement entropies. More importantly, there are no constraints on the subset entanglement entropies as well as the mutual information.

The entanglement entropy for disconnected regions has always been a tough problem in both quantum field theories and quantum many-body systems. For example the entanglement entropy for single intervals in two-dimensional CFT is constrained by symmetries \cite{Calabrese:2004eu}, while the entanglement entropy for multi-intervals depends on the full operator content of the theory \cite{Caraglio:2008pk,Furukawa:2008uk,Calabrese:2009ez,Alba:2009ek,Calabrese:2010he,Coser:2013qda,Ruggiero:2018hyl}. The bipartite correlations $J_{ij}$ may not be enough to characterize the entanglement entropies for disconnected regions. However, the bipartite correlations could be enough to determine the entanglement entropies for connected regions. An interesting fact we find to support this claim is that, for example in the circle lattice with $N$ sites, the number of subset intervals that are connected is $N(N-1)$. Furthermore, if we consider the system to be in a pure state then the number of nonzero entanglement entropies for these intervals becomes $N(N-1)/2$, which exactly matches with the number of $J_{ij}$. If we only apply the normalization requirement for these connected regions, the equations \eqref{IAB2} could be compatible thus have a unique solution. Though we donot have a similar argument for the higher dimensional cases, we still would like to conjecture the claim to be valid.

The \textit{PEE proposal} \eqref{ECproposal} is a much more successful candidate for PEE than the mutual information because it only involves entanglement entropies of connected subsets. Here we would like to propose that \textit{the normalization requirement should only apply to connected regions}.

\section{Partial entanglement entropy in Poincar\'e invariant theories}
\subsection{The general formula for PEE}
In this section we would like to discuss field theories invariant under Poincar\'e symmetries to show how the physical requirements uniquely determines the PEE. This could be achieved following the discussion of Casini and Huerta \cite{Casini:2008wt} on the extensive (or additive) mutual information (EMI). Let us consider a $d$-dimensional quantum field theory where $A$ and $B$ are any two non-intersecting ($d-1$)-dimensional sub-regions. Following the discussion in the previous section, we only consider $A$ and $B$ to be connected regions. As we have shown the requirement of additivity and the invariance under permutation indicate that, the PEE should satisfy the following formula
\begin{align}
\mathcal{I}(A,B)=\int_{A}d\vec{\sigma}_{\textbf{x}}\int_{B}d\vec{\sigma}_{\textbf{y}} \,J(\textbf{x},\textbf{y})\,,
\end{align}
where  $\textbf{x}$ ($\textbf{y}$) represents points in $A$ ($B$), $\vec{\sigma}_{\textbf{x}}$ ($\vec{\sigma}_{\textbf{y}}$) is any infinitesimal subset of  $A$ ($B$) at $\textbf{x}$ ($\textbf{y}$) with a unit normal vector in spacetime. Note that in relativistic theories, the regions are not confined on a time slice. They can be any spacelike surfaces in spacetime, hence we need to include the information of the normal direction at every point. The requirement of causality (requirement-2) claims that for any two regions $A'$ and $B'$ that have the same boundaries as $A$ and $B$, we should have
\begin{align}
 \mathcal{I}(A',B')=\mathcal{I}(A,B)\,.
 \end{align}
This is because any transformation of a region with its boundary fixed can be achieved by a local unitary transformation confined in that region. This constrains the formula \eqref{IAB1} further to be
\begin{align}
\mathcal{I}(A,B)=\int_{A}d\sigma_{\textbf{x}}{}^{\mu}\int_{B}d\sigma_{\textbf{y}}{}^{\nu}\, J_{\mu\nu}(\textbf{x},\textbf{y})\,,
\end{align}
where $\sigma_{\textbf{x}}{}^{\mu}$ is the vector component of the vector $\vec{\sigma}_{\textbf{x}}$ and $J_{\mu\nu}(\textbf{x},\textbf{y})$ is a symmetrical conserved current that satisfies $\partial_{\mu}J^{\mu\nu}(\textbf{x},\textbf{y})=0$. This can be understand by the fact that the flux of a conserved current that passes through a region is invariant under any fluctuation of the region with its boundary fixed. The Poincar\'e invariance furthermore indicates that \cite{Casini:2008wt}
\begin{align}
 J^{\mu\nu}(\textbf{x},\textbf{y})=\frac{(\textbf{x}-\textbf{y})^{\mu}(\textbf{x}-\textbf{y})^{\nu}}{(\textbf{x}-\textbf{y})^{2 d}}G(l)-\frac{g^{\mu\nu}}{(\textbf{x}-\textbf{y})^{2(d-1)}}F(l)\,,
\end{align}
where $l=|\textbf{x}-\textbf{y}|$ is the distance between $\textbf{x}$ and $\textbf{y}$, $F$ and $G$ are two dimensionless functions of $l$. The conservation of $J^{\mu\nu}$ indicates
\begin{align}\label{Cfunction1}
[G(l)-F(l)]'=-(d-1)\frac{2F(l)-G(l)}{l}\,.
\end{align}
The requirement of positivity implies that for any time-like vectors $\vec{\sigma}_{\textbf{x}}$ and $\vec{\sigma}_{\textbf{y}}$, we should have
\begin{align}
\sigma_{\textbf{x}}{}^{\mu}\sigma_{\textbf{y}}{}^{\nu} J_{\mu\nu}(\textbf{x},\textbf{y})\ge 0\,.
\end{align}
This furthermore implies that, 
\begin{align}
2F(l)\geq G(l)\geq 0\,.
\end{align}

Define $C(l)=G(l)-F(l)$, then according to \eqref{Cfunction1} we have
\begin{align}
C'(l)\leq 0\,,
\end{align}
which implies $C(l)$  is always deceasing under the RG flow, hence can be considered as a $c$-function \cite{Zamolodchikov:1986gt,Cardy:1988cwa}. For theories with an infrared fixed point, we have
\begin{align}
 C(l)\geq 0\,,
 \end{align} 
for any $l$. According to \eqref{Cfunction1} we can also write
\begin{align}
F(l)=-\frac{lC'(l)}{d-1}+C(l)\,,\qquad G(l)=-\frac{lC'(l)}{d-1}+2 C(l)\,.
\end{align}
Then it is convenient to define another function $H(l)$ by
\begin{align}
C(l)=(d-1) l^{2 d-3}H'(l)\,.\,
\end{align}
Thus
\begin{align}
J_{\mu\nu}(l)=-\partial_{\mu}\partial_{\nu}H(l)+g_{\mu\nu}\partial_{\alpha}\partial^{\alpha}H(l)\,.
\end{align}
 At last, after we applied the Stokes' theorem we arrive at the following formula for PEE
\begin{align}\label{generalf}
\mathcal{I}(A,B)=\int_{\partial A}\int_{\partial B}d\vec{\eta}_{\textbf{x}} \cdot d\vec{\eta}_{\textbf{y}}\,H(|\textbf{x}-\textbf{y}|)\,,
\end{align}
where $\vec{\eta}_{\textbf{x}}$ and $\vec{\eta}_{\textbf{y}}$ are the infinitesimal subsets on the boundaries $\partial A$ and $\partial B$ with an outward pointing direction in the system and normal to $\partial A$ and $\partial{B}$. The dot means the contraction between vectors. The above equation gives the general formula for the PEE in Poincar\'e invariant field theories. $C(l)$ determines the bipartite correlations of the theory and should depend on other details of the theory. It should be unique when the theory is given, thus determines an unique PEE or entanglement contour.

The requirement of normalization requires the entanglement entropies to be recovered from PEE, i.e.
\begin{align}\label{normalization}
S_{A}=\mathcal{I}(A,\bar{A})\,.
\end{align}
If this hold, then following the requirement of positivity the requirement of upper bound for PEE is automatically satisfied,
\begin{align}
\mathcal{I}(A,\bar{A})>\mathcal{I}(A,B)\,,
\end{align}
because $B\subset\bar{ A}$.

Things become much more determined in the case of conformal field theories. Since $C(l)$ is a $c$-function, it should be a constant in CFTs. Let us define $C(l)=2C_d (d-1)(d-2)$, then we have
\begin{align}\label{Hcft}
H(|\textbf{x}-\textbf{y}|)=-\frac{C_{d}}{|\textbf{x}-\textbf{y}|^{2 d-4}}\,,\qquad d>2\,,
\end{align}
where $C_{d}$ is a constant that depend on $d$. When $d=2$, $H(|\textbf{x}-\textbf{y}|)$ just gives (with a minus sign) the entanglement entropy for the single interval with the end points being $\textbf{x}$ and $\textbf{y}$.

Before going on we would like to comment on the physical interpretation of \eqref{generalf}. The authors of Ref.\cite{Casini:2008wt} interpreted the formula \eqref{generalf} as an extensive (additive) mutual information (EMI)\footnote{The motivation of Ref.\cite{Casini:2008wt} came from the entanglement entropy for multi-intervals in 2-deminsional free massless fermions \cite{Casini:2005rm} (also see Ref.\cite{Casini:2004bw,Calabrese:2004eu,Hubeny:2007re} for similar results), which indicates that the mutual information is additive.}, which is restricted to the special theories \footnote{ Nevertheless, no specific field theories with additive mutual information were known except the free massless fermions in two dimensions \cite{Casini:2005rm}.} where the mutual information is additive (or EMI models). However, in general the mutual information is not additive hence their results seem to be much less generic. Actually we did not use the definition \eqref{dmutual} for mutual information in the derivation, so any quantity that satisfies the requirements-1,2,3 and is symmetrical under the permutation \eqref{permutation} should be given by Eq.\eqref{generalf}. The more natural interpretation for the formula \eqref{generalf} can come from the PEE. And the formula \eqref{generalf} gives the PEE for general Poincar\'e invariant theories in general dimensions.

\subsection{The requirement of normalization}
In the previous subsection we derived a general formula \eqref{generalf} for the PEE in Poincar\'e invariant theories following the physical requirements. Nevertheless, it is too soon to say the formula \eqref{generalf} is the unique solution to all the requirements because we have not used the requirement of normalization and the upper bound. Since the upper bound follows the normalization requirement, we only need to test \eqref{normalization}. The testing is clear for a few-body system as the entanglement entropy is finite, thus can be calculated exactly. For quantum field theories \eqref{normalization} is very subtle because the entanglement entropies usually diverges hence we need to compare between infinite quantities. In CFTs, the PEE seems to be determined up to a single coefficient $C_d$. Note that $C_d$ should be a parameter of the theory thus not depend on the choice of $A$. One may expect that \eqref{normalization} can be satisfied by properly choosing the coefficient $C_{d}$. The PEE  $\mathcal{I}(A,\bar{A})$ is divergent as $l$ vanishes when $\textbf{x}$ and $\textbf{y}$ overlap. In order to test \eqref{normalization}, certain prescriptions are needed to prevent divergence on both sides. 

Some of the calculations of $\mathcal{I}(A,\bar{A})$ for CFTs have been carried out \cite{Casini:2008wt,Swingle:2010jz,Bueno:2015rda,Bueno:2015qya,Bueno:2019mex} using \eqref{generalf}. In these cases the cutoff is given by an uniform distant cutoff between the boundaries of $A$ and $\bar{A}$. Nevertheless, the naive comparison between these results and the known entanglement entropies calculated by replica trick or holographic method \cite{Ryu:2006bv,Ryu:2006ef}, which are regulated by a cutoff at a scale, shows the normalization requirement \eqref{normalization} cannot exactly hold in general. Firstly, for a given CFT, the coefficient $C_d$ we get from imposing the requirement \eqref{normalization} can depend on the choice of $\mathcal{A}$, hence not a parameter determined by the theory. Secondly, consider several different CFTs in $d=3$, the single constant $C_{3}$ is not enough to characterize the difference between the entanglement entropies arising from sharp corners on $\partial\mathcal{A}$ \cite{Casini:2008wt}. Finally, for theories in even dimensions $d\geq 4$, there are two or more trace-anomaly coefficients. In general the entanglement entropy of a region will depend on all of the trace-anomaly coefficients \cite{Solodukhin:2008dh}, rather than a single one, even in a flat background. This contradict with our expectation that the entanglement entropy in CFTs is determined up to a single constant.

The above observations deviate from what we expected and seem to indicate that the requirement of normalization \eqref{normalization} cannot be satisfied in general, hence the physical requirements are too strong to have a solution. This drives the concept of the PEE into a big problem! Nevertheless we would like to point out that, the matching \eqref{normalization} between $\mathcal{I}(A,\bar{A})$ and the entanglement entropies regulated by a scale cutoff is quite subtle. Because, unlike the entanglement entropies regulated at the scale, $\mathcal{I}(A,\bar{A})$ is regulated by ignoring certain local contributions near the entangling surface $\partial A$. The typical way to do the regulation is in the following. Firstly we consider some region $A'\subset A$ with its boundary $\partial\mathcal{A}'$ not intersecting with $\partial A$. Then we calculate the PEE $\mathcal{I}(A',\bar{A})$ which is finite. At last we let $\partial A'$ approach $\partial A$ to get a regulated $\mathcal{I}(A,\bar{A})$. We call this a \textit{local regulation}, which can be achieved in an infinite number of ways.

For example, let us regulate $\mathcal{I}(A,\bar{A})$ with a single infinitesimal parameter $\epsilon$. We denote the points on $\partial A$ as $\textbf{x}$ and the points on $\partial A'$ as $\textbf{y}$. We can either require any one of the spacial coordinate to satisfy $|x_i-y_i|\geq \epsilon$ or require $|\textbf{x}-\textbf{y}|\geq \epsilon$. The regulated $\mathcal{I}(A,\bar{A})$ will differ with the regulation schemes we choose. So the claim that for a given region $A$ its entanglement entropies in different CFTs is determined only up to a single constant is too restrictive and in general not true. Other information that can affect the entanglement entropy would enter the formula \eqref{generalf} through the schemes of the local regulation. Unlike the scale regulation, to specify the local regulation we usually need more than one parameter.

In summary there is no reason to expect $\mathcal{I}(A,\bar{A})$ to exactly recover the entanglement entropies regulated at a scale. Then we want to ask: \textit{Is $\mathcal{I}(A,\bar{A})$ really the entanglement entropy}? A primitive answer to this question is to search for cases where \eqref{normalization} does hold in some way. The only cases that \eqref{normalization} may hold are the \textit{quasi-one-dimensional} cases. Due to the symmetries it is natural to take a uniform distant cutoff $\epsilon$ between $\partial A'$ and $\partial A$, thus the regulation scheme also respects the symmetry. Then it is possible to match the $\mathcal{I}(A,\bar{A})$ with the entanglement entropy $S_{A}$, which is regulated by the scale cutoff $\delta$, by imposing a relation between $\epsilon$ and $\delta$.

Let us consider the entanglement entropy for a disk $A$ with radius $R$ in holographic CFT$_3$. In this case the entanglement entropy can be calculated holographically using the Ryu-Takayanagi (RT) formula \cite{Ryu:2006bv,Ryu:2006ef}. More explicitly the holographic entanglement entropy for a region $A$ is given by the area of the bulk minimal surface (or the RT surface) that is anchored on the boundary of $A$. The area of the minimal surface can be regulated by taking a infinitesimal cutoff $\delta$ at the asymptotic boundary hence gives the following entanglement entropy
\begin{align}\label{SEEdisk}
S_{A}=\frac{c}{6}\frac{2\pi  R}{\delta }-\frac{c}{3} \pi  +\mathcal{O}(\delta)\,.
\end{align} 
On the field theory side $\delta$ is the scale cutoff and $c$ is the central charge. Then we try to recover the above entanglement entropy using the PEE \eqref{generalf}. We consider two co-centric circles with radius $R_{\pm}$ ($R_+\ge R \ge R_-$) which partition the systems into three regions: a disk $A'$ with $r\leq R_-$, the region $\bar{A}$ with $r\geq R_+$ and an annulus $B$ with $R_-<r<R_+$.  When $R_{-}$ approaches $R_+$, we may expect that the PEE $\mathcal{I}(A',\bar{A})$ will recover the entanglement entropy of the disk in some way. Let us define a parameter $\alpha$ in the following way
\begin{align}
R=\frac{R_+ + R_-}{2}+\alpha\frac{(R_+ -R_-)}{2}\,,\quad -1\leq\alpha\leq 1\,.
\end{align}
The calculation of $\mathcal{I}(A',\bar{A})$ is explicitly done in the next section (see also Ref. \cite{Casini:2008wt} where it is calculated as an EMI). Here we just quote the result \eqref{Saa2p} and set $d=3$. Then we find
\begin{align}
\mathcal{I}(A',\bar{A})=C_3\frac{2 \pi ^2  R_- ^2}{R_+^2-R_-^2}\,.
\end{align}
In order to match with the holographic result \eqref{SEEdisk}, we choose $C_3=\frac{c}{3\pi}$ and take the limit $R_+ - R_-=\epsilon\to 0$. Then we find
\begin{align}\label{Saca}
\mathcal{I}(A',\bar{A})|_{\epsilon\to 0}=\frac{c}{6}\frac{2\pi  R}{\epsilon }-\frac{c}{6} \pi  (\alpha +2) +\mathcal{O}(\epsilon)\,.
\end{align}
The first term is the standard area term, while the universal term is ambiguous due to the undetermined parameter $\alpha$.

From the PEE point of view, we should set $R_+=R$ and set $A=A'\cup B$. This corresponds to the choice $\alpha=1$. The PEE is the contribution from $A'$ to the entanglement entropy of the disk $S_{A}$. When $A'$ approaches $A$, i.e. $R_-\to R$, then the PEE recovers the entanglement entropy for the disk. In this case the matching between \eqref{Saca} and \eqref{SEEdisk} is then achieved by imposing the fine correspondence $\epsilon=\delta-\frac{\delta^2}{2R}+\mathcal{O}\left(\delta^3\right)$ \cite{Han:2019scu} between the points on $A$ and the points on the corresponding RT surface. Similarly, we can recover the holographic entanglement entropies for static spheres in higher dimensions and for intervals in $d=2$ using this fine correspondence. In $d=2$, the difference between the two cutoffs $\epsilon$ and $\delta$ only affect entanglement entropy at order $\mathcal{O}(\epsilon)$ thus can be ignored. This is not true in $d > 2$

Also another treatment using EMI \cite{Casini:2015woa} is worth mentioning. It was argued that in order to protect the universal term from the UV physics, we should choose $\alpha=0$. It is also obvious that when $\alpha=0$ Eq.\eqref{Saca} exactly matches with Eq.\eqref{SEEdisk} after we replace $\epsilon$ with $\delta$. It is claimed \cite{Casini:2015woa} that this argument can be extend to boundaries with any shape. Nevertheless, this treatment could fail for higher dimensional cases.

For spheres, cylinders or static intervals, if we again choose the regulation scheme that respect the symmetry, there are only one parameters $C_d$ (except the cutoff $\epsilon$) that we can adjust to satisfy the normalization condition \eqref{normalization}. This can not be satisfied if the entanglement entropies on the left hand side depend on more than one conformal anomaly. So far the entanglement entropies for spheres or cylinders (four dimensions) in CFTs with a scale cutoff are carried out through different approaches \cite{Solodukhin:2008dh,Bueno:2015rda}. As expected, they only depend on one conformal anomaly. For the cases that are not \textit{quasi-one-dimensional}, the scheme of the local regulation should at least depend on the space-time curvature near the boundary and the extrinsic geometric of the boundary $\partial A$, which is crucial and not emphasized before. It will be very interesting to explore the relation between \eqref{normalization} and Solodukhin's general formula for four-dimensional CFTs \cite{Solodukhin:2008dh}.

\subsection{Consistency with the PEE proposal}
Our previous discussion indicates that, despite the subtlety of the requirement of normalization, the physical requirements give strong enough constraints to determine the PEE in Poincar\'e invariant field theories. Previously we have shown that, in \textit{quasi-one-dimensional} cases the \textit{PEE proposal} \eqref{ECproposal} is an exact solution for all the requirements in general theories. So the unique solution in Poincar\'e invariant theories should be the \textit{PEE proposal}. This means \eqref{ECproposal} should be consistent with the formula \eqref{generalf}. Furthermore since the \textit{PEE proposal} satisfies the requirement of normalization with no subtlety, the formula \eqref{generalf} should also satisfy \eqref{normalization}. This means $\mathcal{I}(A,\bar{A})$ is indeed the entanglement entropy (at least in the \textit{quasi-one-dimensional} cases), thus answering our previous question. Here we directly prove this equivalence between \eqref{ECproposal} and \eqref{generalf}.

Since PEE reduces to an integration on relevant boundaries, we can write it as a functional of the boundaries with directions,
\begin{align}\label{peeb}
s_{\mathcal{A}}(\mathcal{A}_i)=\mathcal{I}(\mathcal{A}_i,\bar{\mathcal{A}})=\tilde{\mathcal{I}}(\overrightarrow{\partial \mathcal{A}_i},\overrightarrow{\partial \bar{\mathcal{A}}})\,,
\end{align}
where $\overrightarrow{\partial \mathcal{A}_i}$ is defined as the boundary $\partial \mathcal{A}_i$ with an outward-pointing direction. Under this notation, we should have properties such as $\overrightarrow{\partial \mathcal{A}}=-\overrightarrow{\partial \bar{\mathcal{A}}}$ and $\tilde{\mathcal{I}}(\overrightarrow{\partial \mathcal{A}_i},-\overrightarrow{\partial \bar{\mathcal{A}}})=-\tilde{\mathcal{I}}(\overrightarrow{\partial \mathcal{A}_i},\overrightarrow{\partial \bar{\mathcal{A}}})$. 
 For example, consider the cases in Fig.\ref{1}, where $\mathcal{A}$ is partitioned into $\{\mathcal{A}_1, \mathcal{A}_2 ,\mathcal{A}_3\}$. We denote the boundaries of $\mathcal{A}$ as $L_1$ and $L_2$ while the boundaries of $\mathcal{A}_2$ as $l_1$ and $l_2$. Since we need to specify the direction of the boundaries when calculating PEE, it is convenient to define that $L_i$ ($l_i$) points outward from $\mathcal{A}$ ($\mathcal{A}_2$), while $-L_i$ ($-l_i$) points inward. Following this notation we have
\begin{align}
&\overrightarrow{\partial\mathcal{A}_1}:\{L_2,-l_2\},\quad  \overrightarrow{\partial(\mathcal{A}_1\cup\mathcal{A}_2)}:\{L_2,l_1\}\,,
\cr
&\overrightarrow{\partial\mathcal{A}_3}:\{L_1,-l_1\}\,,\quad \overrightarrow{\partial(\mathcal{A}_2\cup \mathcal{A}_3)}:\{L_1,l_2\}\,,
\cr
&\overrightarrow{\partial\bar{\mathcal{A}}}:\{-L_1,-L_2\}\,.
\end{align}
The formula \eqref{generalf} can be written as \eqref{peeb} for short, so we have
\begin{align}\label{leftEC}
s_{\mathcal{A}}(\mathcal{A}_2)=&\tilde{\mathcal{I}}(\overrightarrow{\partial\mathcal{A}_2},\overrightarrow{\partial\bar{\mathcal{A}}})=\tilde{\mathcal{I}}(\{l_1,l_2\},\{-L_1,-L_2\})
\cr
=&-\tilde{\mathcal{I}}(l_1,L_1)-\tilde{\mathcal{I}}(l_1,L_2)
-\tilde{\mathcal{I}}(l_2,L_1)-\tilde{\mathcal{I}}(l_2,L_2)\,.
\end{align}

On the right hand side of Eq.\eqref{ECproposal} the subset entanglement entropies can be calculated via Eq.\eqref{normalization}. For example the entanglement entropy $S_{\mathcal{A}_1}$ is given by
\begin{align}
S_{\mathcal{A}_1}=&\tilde{\mathcal{I}}\left(\{L_2,-l_2\},\{-L_2,l_2\}\right)
\cr
=&\tilde{\mathcal{I}}(L_2,-L_2)+2\tilde{\mathcal{I}}(L_2,l_2)+\tilde{\mathcal{I}}(-l_2,l_2)\,.
\end{align}
Similarly, we have
\begin{align}
S_{\mathcal{A}_3}&=\tilde{\mathcal{I}}(-l_1,l_1)+2\tilde{\mathcal{I}}(L_1,l_1)+\tilde{\mathcal{I}}(L_1,-L_1)\,,
\cr
S_{\mathcal{A}_1\cup\mathcal{A}_2}&=\tilde{\mathcal{I}}(l_1,-l_1)-2\tilde{\mathcal{I}}(l_1,L_2)+\tilde{\mathcal{I}}(L_2,-L_2)\,,
\cr
 S_{\mathcal{A}_2\cup\mathcal{A}_3}&=\tilde{\mathcal{I}}(l_2,-l_2)-2\tilde{\mathcal{I}}(l_2,L_1)+\tilde{\mathcal{I}}(L_1,-L_1)\,.
\end{align}
We then plug the above entanglement entropies into Eq.\eqref{ECproposal}. All the subtle divergent terms cancel and the remaining terms are cutoff independent and exactly match with \eqref{leftEC}. Hence the consistency between the proposal \eqref{ECproposal}  and the formula \eqref{generalf} is justified. The linear combination is absolutely not a mutual information, hence the interpretation of \eqref{generalf} as an extensive mutual information \cite{Casini:2008wt} is in general not true.

If the normalization requirement can be satisfied by any connected regions, we can furthermore generalize the proposal \eqref{ECproposal} to a a generic set up. Given a generic connected $\mathcal{A}$ with outward-pointing boundaries $\{L_j\}$ and a connected subset $\mathcal{A}_{1}$ with outward-pointing boundaries $\{l_i\}$, it is easy to derive that (see Appendix \ref{C}) the PEE $s_{\mathcal{A}}(\mathcal{A}_1)$ can be written as the following linear combination of connected subset entanglement entropies,
\begin{align}\label{ECproposalg}
s_{\mathcal{A}}(\mathcal{A}_1)=&\sum_{outward}\frac{1}{2}\left(S_{\mathcal{A}_{ij}}-S_{l_i}-S_{L_j}\right)
\cr
&-\sum_{inward}\frac{1}{2}\left(S_{\mathcal{A}_{ij}}-S_{l_{i}}-S_{L_{j}}\right)\,.
\end{align}
In the above equation $\mathcal{A}_{ij}$ is the region enclosed by $L_i$ and $l_j$, $S_{l_i}$ $(S_{L_j})$ is the entanglement entropy of the region enclosed by the single boundary $l_i$ ($L_j$). In the first summation $l_i$ points outward $\mathcal{A}_{ij}$ while in the second summation $l_i$ points inward $\mathcal{A}_{ij}$. This generalization \eqref{ECproposalg} is bold and need further investigation. 

Note that like the \textit{PEE proposal}, the generalization \eqref{ECproposalg} also only involves entanglement entropies for connected regions. 

\subsection{Matching with the fine structure analysis}
Another independently developed proposal to construct the entanglement contour function is the fine structure analysis of the entanglement wedge \cite{Wen:2018whg,Han:2019scu}. The strategy is to consider the bulk extension of the boundary modular flow lines, which are two-dimensional surfaces that form a natural slicing of the bulk entanglement wedge. This slicing relates the points on the boundary region $\mathcal{A}$ to the points on the RT surface $\mathcal{E}_{\mathcal{A}}$ by static spacelike geodesics normal to $\mathcal{E}_{\mathcal{A}}$.  Based on this fine relation the contour function for static spherical regions (or intervals) in the vacuum state of holographic CFTs were carried out in Refs.\cite{Wen:2018whg,Han:2019scu} (see also Ref.\cite{Kudler-Flam:2019oru}). It is given by,
 \begin{align}\label{contoursph}
f_{\mathcal{A}}(r)=\frac{c_d}{6}\left(\frac{2R}{R^2-r^2}\right)^{d-1}\,,
\end{align}
where $R$ is the radius of the spherical region $\mathcal{A}$, $d$ is the spacetime dimension, and $c_d=a_{d}^* \frac{2 \Gamma(d/2)}{\pi^{d/2-1}}$ (see Ref.\cite{Myers:2010xs,Myers:2010tj} for the definition of $a_{d}^*$) is a constant related to the A-type central charge. Following Eq.\eqref{a2definition} it is easy to calculate the PEE \cite{Han:2019scu},
\begin{align}\label{Saa2}
s_{\mathcal{A}}(\mathcal{A}_2)=&\frac{c_d}{6}\int_{0}^{R_0} \left(\frac{2R}{R^2-r^2}\right)^{d-1}\Omega_{d-2} r^{d-2}~ dr
\cr
=&\frac{ c_d}{6}  (4\pi z^2) ^{\frac{d-1}{2}} \, _2\tilde{F}_1(\frac{d-1}{2},d-1;\frac{d+1}{2};z^2)\,,
\end{align}
where $\mathcal{A}_2$ is a cocentric sphere with radius $R_0<R$, $ _2\tilde{F}_1\left(a,b,c,x\right)=~_2F_1\left(a,b;c;x\right)/\Gamma(c)$ is the regularized hypergeometric function, $\Omega_{d-2}$ is the volume of the unit $(d-2)$-sphere $\mathbf{S}^{d-2}$ and $z$ is the ratio $z=R_0/R$.

On the other hand, in these cases we can also calculate the PEE using the formula \eqref{generalf} on an infinitely big plane. Plugging \eqref{Hcft} into \eqref{generalf}, we find \cite{Casini:2008wt}
\begin{align}\label{Saa2p}
\mathcal{I}(\bar{\mathcal{A}},\mathcal{A}_2)=&-\int_{\partial A}  \int_{\partial B} d\vec{\eta}_{\textbf{x}} \cdot d\vec{\eta}_{\textbf{y}} \,\frac{C_{d}}{|\textbf{x}-\textbf{y}|^{2d-4}}
\cr
=&C_{d}\Omega_{d-2}\Omega_{d-3} \int^{\pi	}_{0}d\theta\frac{z^{d-2}(\sin\theta)^{d-3}\cos\theta}{\left(1+z^2-2z \cos\theta\right)^{d-2}}\,.
\end{align}
Though it is not easy to write the above integration in a compact form as in Eq.\eqref{Saa2}, one can check that the integration \eqref{Saa2p} coincide \footnote{Note that the PEE \eqref{Saa2} is semi-classical while \eqref{Saa2p} is quantum, which indicates the quantum correction to \eqref{Saa2} is proportional to \eqref{Saa2}.} with \eqref{Saa2} after we properly fix the coefficient $C_d$ that depend on $d$. For example when $d=\{3,4,5\}$ we have
\begin{align}
&s_{\mathcal{A}}(\mathcal{A}_2)\sim \mathcal{I}(\bar{\mathcal{A}},\mathcal{A}_2)\sim \Big\{\frac{z^2}{1- z^2}\,,
 \frac{z^3+z}{\left(z^2-1\right)^2}-\frac{1}{2} \tanh ^{-1}\left(\frac{2 z}{z^2+1}\right)\,,~\frac{ z^4 \left(z^2-3\right)}{ \left(z^2-1\right)^3}\Big\}.
\end{align}

The above contour functions or PEEs can also be reproduced by the \textit{PEE proposal} \eqref{ECproposal} \cite{Han:2019scu}. Note that naively plugging the holographic entanglement entropies for annuli \cite{Fonda:2014cca,Nakaguchi:2014pha} into \eqref{ECproposal} will give a different answer for PEE. More explicitly the PEE vanishes when $R_0$ is smaller than the critical value where the RT surface for the annulus switch between the connected half-torus surface and two disconnected semi-spheres.

\section{Discussion}
In this paper we have explored the concept and properties of the partial entanglement entropy in many aspects. Since it is quite closely related to the previous papers \cite{Wen:2018whg,Kudler-Flam:2019oru,Wen:2019ubu}, it will be useful to clarify what are the new results in this paper.  In the following is a brief summary on the main results of this paper:
\begin{enumerate}
\item Since the entanglement or correlation is mutual we introduce a new physical requirement for PEE, i.e. $\mathcal{I}(A,B)=\mathcal{I}(B,A)$.

\item We show in the \textit{quasi-one-dimensional} cases the \textit{PEE proposal} is a solution to all the known physical requirements (including the new one) in general theories.

\item We suggest not to apply the normalization requirement when evaluating entanglement entropies for disconnected regions via PEE. The \textit{PEE proposal} is not equivalent to the extensive (additive) mutual information \cite{Casini:2008wt,Roy:2019gbi}.

\item We show in Poincar\'e invariant theories the PEE can be uniquely determined by the requirements. The result is consistent with the \textit{PEE proposal} and the fine structure analysis. So the \textit{PEE proposal} is justified in Poincar\'e invariant theories. This is for the first time we have found that the PEE can be uniquely determined. 

\item The subtlety of the normalization requirement is discussed. We clarify that the entanglement entropy approximated by PEE is regulated by excluding the local contributions near the boundary, which is totally different from the familiar entanglement entropies regulated at a scale. Hence, in general we should not expect the exact matching between these two kinds of prescriptions.

\end{enumerate}

One of the important lessons we can learn from our discussion for Poincar\'e invariant field theories is that the PEE $\mathcal{I}(A,\bar{A})$ does give the entanglement entropy. And this claim is valid for general Poincar\'e invariant theories rather than confined to the EMI models \cite{Casini:2008wt,Roy:2019gbi}. We donot expect $\mathcal{I}(A,\bar{A})$ and the entanglement entropies regulated at a scale to exactly match with each other, but we do expect them to respect the same features. This is confirmed by the previous evaluations \cite{Casini:2008wt,Swingle:2010jz,Bueno:2015rda,Bueno:2015qya,Bueno:2019mex} of entanglement entropies (locally regulated by a uniform distance at the boundary) approximated by PEE (or EMI) in various dimensions and various theories. These evaluations give support to our claim that the bipartite correlations could be enough to determine the entanglement entropies for connected regions. Also \eqref{normalization} can reproduce the known universal results for generic CFTs, for example the corner entanglement entropy in three dimensions at the smooth limit \cite{Bueno:2015qya,Bueno:2015rda}, and the entanglement entropy for a conical entangling surface in four dimensions \cite{Bueno:2019mex}.

There are also attempts to calculate the entanglement entropies for disconnected regions using \eqref{generalf}.  For example, naively taking an uniform cutoff for all the endpoints of multi-intervals in CFT$_2$ will give \cite{Berthiere:2019lks,Roy:2019gbi} the results of Refs. \cite{Casini:2005rm,Casini:2004bw,Calabrese:2004eu,Hubeny:2007re}, which are not correct in general. More explicitly it only captures part of the entanglement entropy and misses the term that depends on the harmonic ratio of the entangling point \cite{Caraglio:2008pk,Furukawa:2008uk,Calabrese:2009ez,Alba:2009ek,Calabrese:2010he,Coser:2013qda,Ruggiero:2018hyl}. Bipartite correlation may not be enough to determine the entanglement among three or more regions. How to generalize the concept of PEE to the cases that involve more than two connected regions is an important future direction.

Our proof of the uniqueness of the PEE in Poincar\'e invariant theories relies on the symmetry. It will be important to explore the uniqueness of the PEE in more general field theories and quantum many-body systems. The new requirement we have introduced may help us rule out some candidates to construct the entanglement contour.

The PEE is a new concept in quantum information. Its potential to help us better understand the entanglement structure of a quantum system should be further investigated. We encourage people to apply the \textit{PEE proposal} \eqref{ECproposal} to few-body systems or lattice models in condense matter theories (especially in $d=2$) to calculate the entanglement contour function. Then it will be interesting to compare it with other measurements. Here we would like to mention a recent work \cite{MacCormack:2020auw}, where the entanglement contour calculated by the \textit{PEE proposal} has been investigated for two distinct non-thermalizing phases: many-body localization (MBL) and the random singlet phase (RSP). Novel properties of entanglement spreading are revealed by the entanglement contour, which goes beyond the measure of the out-of-time-ordered correlator (OTOC).  For example, they found a logarithmic light cone of entanglement spreading in MBL from the entanglement contour after a global quench, which was similar but not identical to the logarithmic light cone seen for the OTOC. Also in the RSP, the entanglement contour yielded a novel power-law light cone, despite trivial spreading of the OTOC in that system.

\section*{Acknowledgments}
The author would like to thank Chong-Sun Chu, Jonah Kudler-Flam, Rong-xin Miao, Tatsuma Nishioka, William Witczak-Krempa, Zhuo-Yu Xian,  and Gang Yang for helpful discussions. Especially I would like to thank Chen-Te Ma for a careful reading of the manuscript and Horacio Casini for pointing out their work on mutual information and $c$-functions. I thank the Yukawa Institute for Theoretical Physics at Kyoto University. Discussions during the workshop YITP-T-19-03 ``Quantum Information and String Theory 2019'' were useful to complete this work. This work is supported by the NSFC Grant No.11805109 and the start-up funding of the Southeast University.

\appendix

\section{Partial entanglement entropy for general partitions}\label{C}

\begin{figure}[h] 
   \centering
      \includegraphics[width=0.3\textwidth]{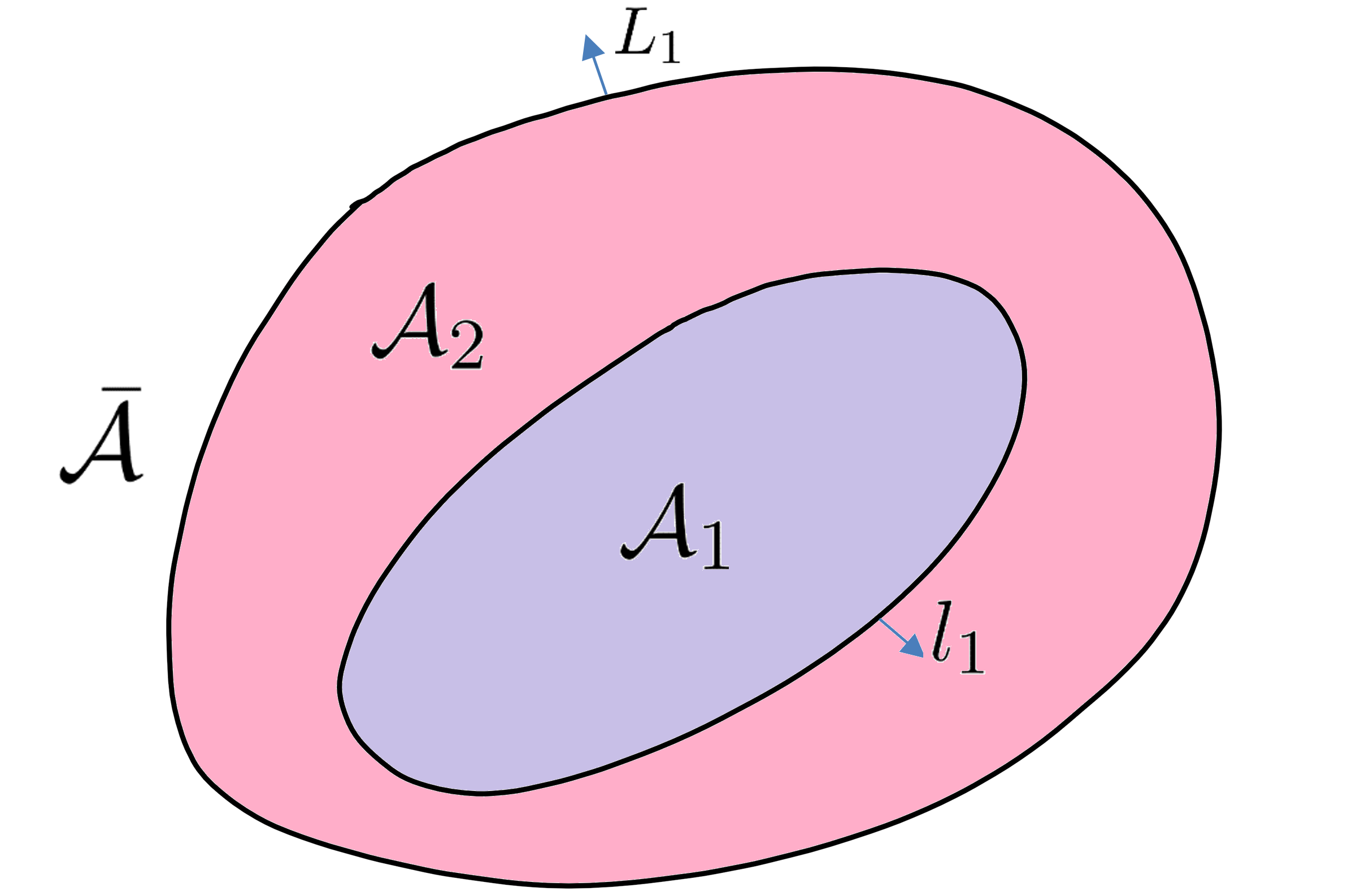}\quad \includegraphics[width=0.3\textwidth]{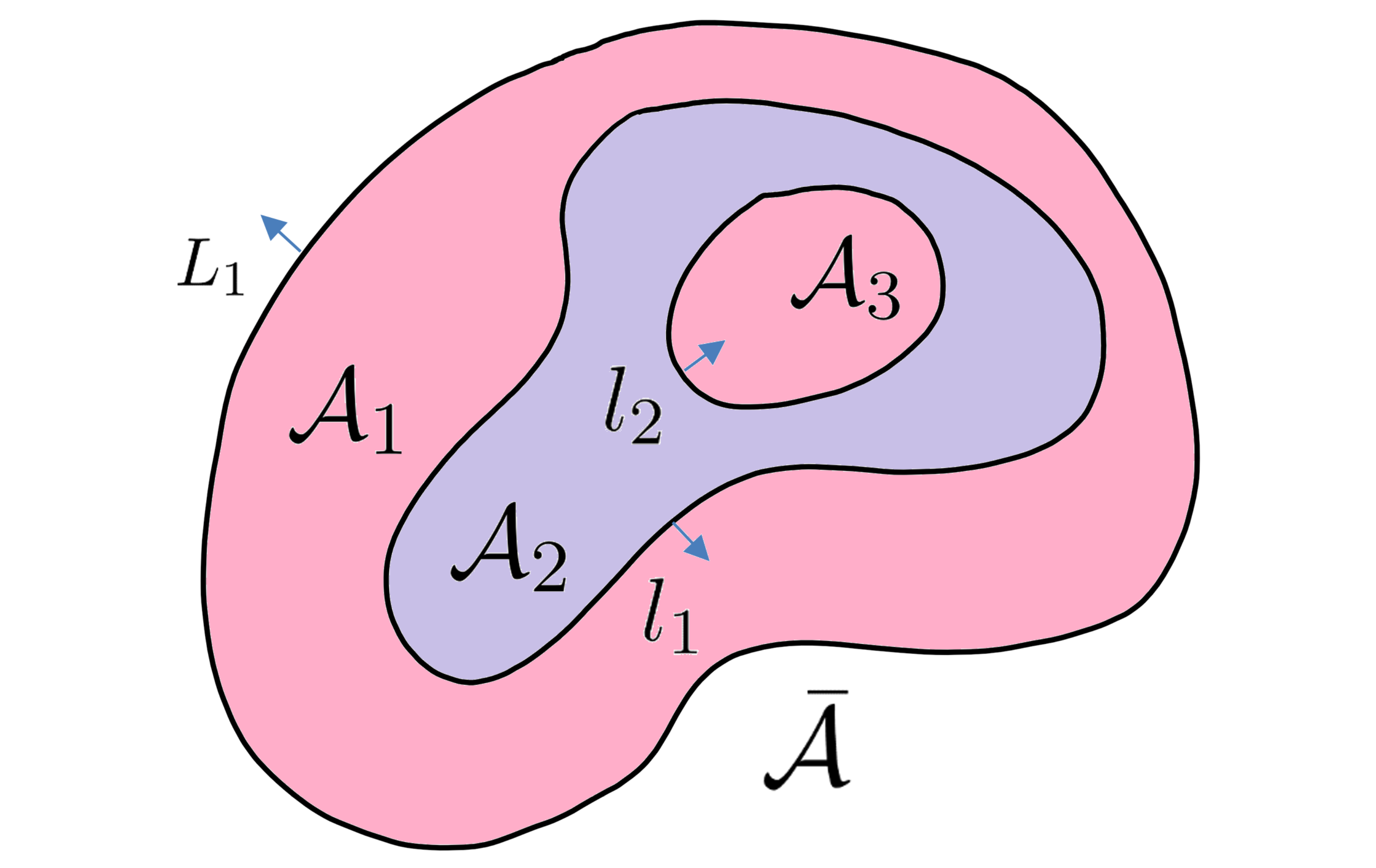} \quad  \includegraphics[width=0.3\textwidth]{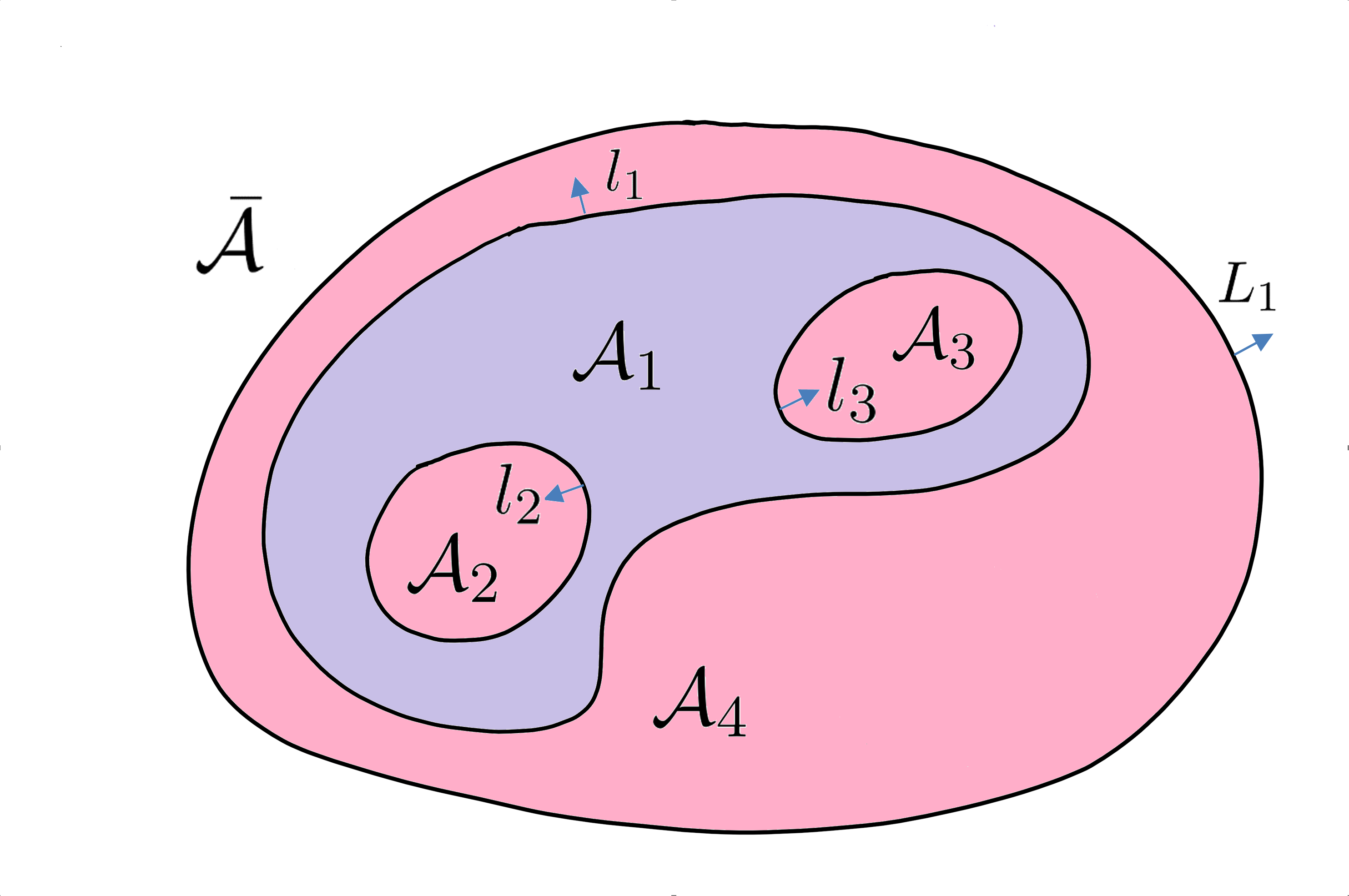}
 \caption{The colored region is $\mathcal{A}$ and the blue region is the subset $\mathcal{A}_i$ whose PEE we are considering. We calculate $s_{\mathcal{A}}(\mathcal{A}_i)$, so $l_i$ ($L_j$) points outward $\mathcal{A}_1$ ($\mathcal{A}$). From left to right the figures correspond to case 1, case 2 and case 3.
\label{45}}
\end{figure}

We generalize the \textit{PEE proposal} \eqref{ECproposal} to a generic partition. We consider a generic region $\mathcal{A}$ with $m$ outward-pointing boundaries $L_j\,(1\leq j \leq m)$, and a subset $\mathcal{A}_1$ with $n$ outward-pointing boundaries $l_i\,(1\leq i \leq n)$ (see, for example, Fig.\ref{45}). Then the PEE $s_{\mathcal{A}}(\mathcal{A}_1)=\mathcal{I}(\mathcal{A}_1,\bar{\mathcal{A}})$ is given by
\begin{align}
s_{\mathcal{A}}(\mathcal{A}_1)=&\tilde{\mathcal{I}}(\{l_1,\cdots , l_n\},\{-L_1,\cdots, -L_m\})
\cr
=&\sum_{i=1}^{n}\sum_{j=1}^{m}\tilde{\mathcal{I}}(l_i,-L_j)\,.
\end{align}
We try to write the summation as a linear combination of entanglement entropies. We denote the region enclosed by the two boundaries $l_i$ and $L_j$ as $\mathcal{A}_{ij}$. Since $\mathcal{A}_{ij}$ is always inside $\mathcal{A}$, so $L_j$ is also the outward-pointing boundary of $\mathcal{A}_{ij}$. However $l_i$ could be either the outward-pointing or inward-pointing boundary of $\mathcal{A}_{ij}$. For example, in the right figure of Fig. \ref{45}, $l_1$ is the inward-pointing boundary of $\mathcal{A}_{1,j}$, while $l_2$ ($l_3$) is the outward-pointing boundary of $\mathcal{A}_{2,j}$ ($\mathcal{A}_{3,j}$).
When $l_i$ is the outward-pointing boundary of $\mathcal{A}_{ij}$, according to \eqref{normalization} we have
\begin{align}
&S_{\mathcal{A}_{ij}}=\tilde{\mathcal{I}}(l_i,-l_i)+\tilde{\mathcal{I}}(L_j,-L_j)+2\tilde{\mathcal{I}}(l_i,-L_j)\,,
\\\label{Iij1}
&\tilde{\mathcal{I}}(l_i,-L_j)=\frac{1}{2}\left(S_{\mathcal{A}_{ij}}-S_{l_i}-S_{L_j}\right)\,.
\end{align}
$S_{l_i}$ $(S_{L_j})$ is the entanglement entropy of the region enclosed by $l_i$ ($L_j$). 
When $l_i$ is the inward-pointing boundary of $\mathcal{A}_{ij}$, we should have
\begin{align}
&S_{\mathcal{A}_{ij}}=\tilde{\mathcal{I}}(l_i,-l_i)+\tilde{\mathcal{I}}(L_j,-L_j)+2\tilde{\mathcal{I}}(l_i,L_j)\,,
\\\label{Iij2}
&\tilde{\mathcal{I}}(l_i,-L_j)=\frac{1}{2}\left(S_{l_i}+S_{L_j}-S_{\mathcal{A}_{ij}}\right)\,.
\end{align}

Then we conclude that, given a generic $\mathcal{A}$ with outward-pointing boundaries $L_j$ and a subset $\mathcal{A}_{1}$ with outward-pointing boundaries $l_i$, the PEE $s_{\mathcal{A}}(\mathcal{A}_1)$ should be given by
\begin{align}\label{PEE}
s_{\mathcal{A}}(\mathcal{A}_1)=&\sum_{outward}\frac{1}{2}\left(S_{\mathcal{A}_{ij}}-S_{l_i}-S_{L_j}\right)
\cr
&- \sum_{inward}\frac{1}{2}\left(S_{\mathcal{A}_{ij}}-S_{l_{i}}-S_{L_{j}}\right)\,.
\end{align}

In the following we apply Eq.\eqref{PEE} to the three cases in Fig. \ref{45}.
\begin{itemize}
\item\textbf{Case 1:} The subset $\mathcal{A}_1$ partition the $\mathcal{A}$ into two parts and $l_1$ points inward $\mathcal{A}_{11}=\mathcal{A}_1$. So we have
\begin{align}
2s_{\mathcal{A}}(\mathcal{A}_1)=S_{\mathcal{A}}+S_{\mathcal{A}_1}-S_{\mathcal{A}_2}=I(\bar{\mathcal{A}},\mathcal{A}_1)\,.
\end{align}
In this case $2 s_{\mathcal{A}}(\mathcal{A}_1)$ is a mutual information. If we further divide $\mathcal{A}_1$ into subsets $\mathcal{A}_1^{j}$ with the same topology of a sphere, then $2 s_{\mathcal{A}}(\mathcal{A}_i)$ is also the mutual information $I(\bar{\mathcal{A}},\mathcal{A}_{1}^{i})$. Because PEE is additive, thus $I(\bar{\mathcal{A}},\mathcal{A}_1)=\sum_{i}I(\bar{\mathcal{A}},\mathcal{A}_{1}^{i})$ looks additive. However, when any of the $\mathcal{A}_{1}^{i}$ has other topologies (such as an annulus as in the previous case), this additivity for mutual information breaks down.

\item \textbf{Case 2:} We calculate $s_{\mathcal{A}}(\mathcal{A}_2)$. It is easy to see that $S_{\mathcal{A}_{11}}=S_{\mathcal{A}_1}, S_{L_1}=S_{\mathcal{A}}, S_{l_1}=S_{\mathcal{A}_2\cup\mathcal{A}_3}, S_{l_2}=S_{\mathcal{A}_3}$, $-l_1$ points inward $\mathcal{A}_{11}$ and $l_2$ points outward $\mathcal{A}_{21}$. Then according to \eqref{PEE} we find
\begin{align}
2 s_{\mathcal{A}}(\mathcal{A}_2)
=&S_{\mathcal{A}_1\cup\mathcal{A}_2}+S_{\mathcal{A}_2\cup \mathcal{A}_3}-S_{\mathcal{A}_1}-S_{\mathcal{A}_3}\,,
\end{align}
which is absolutely not the mutual information $I(\bar{\mathcal{A}},\mathcal{A}_2)$.

\item\textbf{Case 3:} In this case we find
\begin{align}
2s_{\mathcal{A}}(\mathcal{A}_1)= &S_{\mathcal{A}_1\cup\mathcal{A}_3\cup\mathcal{A}_4}+S_{\mathcal{A}_1\cup\mathcal{A}_2\cup\mathcal{A}_4}+S_{\mathcal{A}_1\cup\mathcal{A}_{2}\cup\mathcal{A}_{3}}
\cr
&
-S_{\mathcal{A}_4}-S_{\mathcal{A}_{2}}-S_{\mathcal{A}_{3}}-S_{\mathcal{A}}\,.
\end{align}

\end{itemize}


\bibliographystyle{JHEP}
 \bibliography{lmbib}

\end{document}